\documentclass[aps,prb,preprint,showpacs,floatfix,superscriptaddress]{revtex4-1}
\usepackage{epsfig,graphicx,amsmath,txfonts}
\usepackage{sidecap}
\newcommand*{\et}[0]{\textit{et~al.}}
\newcommand*{\fig}[1]{Figure~\ref{fig:#1}}
\newcommand*{\tab}[1]{Table~\ref{tab:#1}}

\newcommand*{\TB}[0]{(10\bar12)}
\newcommand*{\SF}[0]{(10\bar10)}

\begin{document}

\title{Interaction of oxygen interstitials with lattice faults in Ti}
\author{M. Ghazisaeidi}
\email{ghazisaeidi.1@osu.edu}
\affiliation{Department of Materials Science and Engineering, Ohio State University, Columbus, OH 43210, USA}
\author{D. R. Trinkle}
\affiliation{Department of Materials Science and Engineering, Univ. Illinois, Urbana-Champaign, Urbana, IL 61801, USA}

\begin{abstract}
Oxygen greatly affects the mechanical properties of titanium. In addition, dislocations and twin boundaries influence the plastics deformation of hcp metals. As part of a systematic study of defects interactions in Ti, we investigate the interactions of oxygen with $\TB$ twin boundary and $\SF$ prism plane stacking fault. The energetics of four interstitial sites in the twin geometry are compared with the bulk octahedral site. We show that two of these sites located at the twin boundary are more attractive to oxygen than bulk while the sites away from the boundary are repulsive. Moreover, we study the interaction of oxygen with the prismatic stacking fault to approximate oxygen-dislocation interaction. We show that oxygen increases the stacking fault energy and therefore is repelled by the faulted geometry and consequently a dislocation core.
\end{abstract}

\maketitle

\section{Introduction}
Titanium has excellent strength to weight ratio, high melting temperature and good corrosion resistance which render it suitable for light-weight, high temperature applications\cite{Long:1998fk}. The tensile and fatigue strength of Ti increase with oxygen content at the cost of compromising ductility and toughness~\cite{ref:wasz,Titanium2007}. These changes in mechanical properties are attributed to the interstitial oxygen atoms impeding dislocation motion and suppressing low temperature twinning~\cite{ref:conrad}. Understanding mechanisms underlying this effect of oxygen is thus important from both scientific and industrial points of view. 

A recent model developed by Oberson \et~\cite{ref:oberson} suggests that twinning in $\alpha$-Ti is controlled by diffusion of oxygen. This model is based on the assumption that oxygen would tend to escape away from the twin boundaries due to annihilation of the bulk octahedral sites for interstitials. However, being purely crystallographic, this model does not account for the details of atomic-scale interaction between the oxygen atom and its surrounding Ti neighbors. Despite numerous DFT studies of the effect of oxygen on various properties in titanium\cite{Hennig2005,Wu:2011fk,Burton:2012uq,Wu2013a}, there has yet to be a first-principles study of oxygen interacting with lattice faults in titanium. Here, we investigate such interactions with density functional theory (DFT) calculations and show that the oxygen atom induces significant relaxations on its nearest neighbor Ti atoms, thereby creating new interstitial sites at and around the twin boundary. Our results show that the interstitial sites located at the twin boundary are in fact more energetically favorable for the oxygen atom compared to bulk octahedral sites.

In addition, quantifying the interaction between oxygen interstitials and dislocations in Ti is instrumental to understanding the mechanisms of the observed strengthening effect. Previously, we computed the core geometry of screw dislocations in Ti using DFT and Green's function boundary conditions~\cite{ref:screw}. However, calculation of dislocation/oxygen interaction energy requires extra care: Periodic boundary conditions create an infinite array of oxygen atoms along the dislocation line, rather than a single, isolated oxygen atom. To approximate the dilute limit, periodicity along the dislocation threading direction should be increased to a value where the interaction between the solute and its neighboring images is negligible, or at least the \textit{difference} in interaction is negligible compared to bulk. Selecting a suitable distance requires calculations of oxygen-oxygen interactions at various separations in bulk Ti. Here we present these calculations and show that there is a significant repulsive interaction between the oxygen interstitial and its image at one lattice parameter distance. We find that to get reasonably small interactions, the distance between periodic images of oxygen atom should be at least twice the lattice parameter if the oxygen is in a bulk octahedral site and even more if it is in a defective site. Therefore, to model the interaction energy between oxygen and the isolated screw dislocation reported in~\cite{ref:screw}, the supercell must contain at least 1864 atoms. This is extremely expensive for DFT calculations. Instead, we compute the interaction energy of the oxygen atom with prismatic stacking faults in Ti as a surrogate for the dislocation core and show that oxygen increases the stacking fault energy. Overall, the present study provides various pieces of quantitative information on the interaction of oxygen interstitials with defects in Ti and is a first step towards understanding the effect of oxygen on slip and twinning in Ti.

\section{computational method}
\textit{Ab initio} calculations of oxygen in Ti are performed with \textsc{vasp}~\cite{ref:vasp1,ref:vasp2}, a plane wave based density functional code using projected 
augmented wave (PAW) method within generalized gradient approximation (GGA)~\cite{ref:gga}. Ti 4\textit{s} and 3\textit{d} electrons are considered as valence electrons. A planewave energy cut-off of 500 eV is used throughout the calculations. A k-points mesh of 16$\times$16$\times$12 is used for a Ti unit cell and is adjusted for each geometry accordingly.~\tab{compare} compares the lattice and elastic constants and planar fault energies obtained from PAW, ultra-soft pseudo potentials (USPP) with \textit{p} electrons treated as valence and experiments. Lattice parameters and twin boundary energies from PAW and USPP are almost identical. The elastic constants differ within 12$\%$ and the prismatic stacking fault energy is approximately 20$\%$ higher for USPP. Both PAW and USPP agree well with experiments.    

\begin{table}[!ht]
\caption{Comparison of lattice and elastic constants and planar fault energies for PAW with USPP and experiments. The elastic constants are in GPa, and the two fault energies---$\TB$ twin boundary (TB) and $\SF$ prismatic stacking fault (SF)---are given in J/m$^2$.}
\label{tab:compare}

\centering
\begin{ruledtabular}
\begin{tabular}{cccccccccc}
       & \multicolumn{2}{c} {lattice constants} & \multicolumn{5}{c} {elastic constants} & \multicolumn{2}{c} {fault energies} \\
         & $a$ (\AA) & $c/a $ & $C_{11} $& $C_{33}$& $C_{44}$& $C_{13}$&$C_{12}$ & TB   & SF  \\ 
\hline
PAW  & 2.9197 & 1.581 & 169 & 189 & 37 & 84 &97 & 0.296& 0.219 \\
USPP  &2.9486 & 1.580 &164 & 190 & 42 & 75 & 89 & 0.300 & 0.264\\
Experiment & 2.95 & 1.587 & 176 & 191 & 51 & 68 & 87 & --- &  0.150
\end{tabular}
\end{ruledtabular}
\end{table}

\section{Twin boundary}
\begin{figure}
 \begin{center}
   \includegraphics[height=0.5\textwidth]{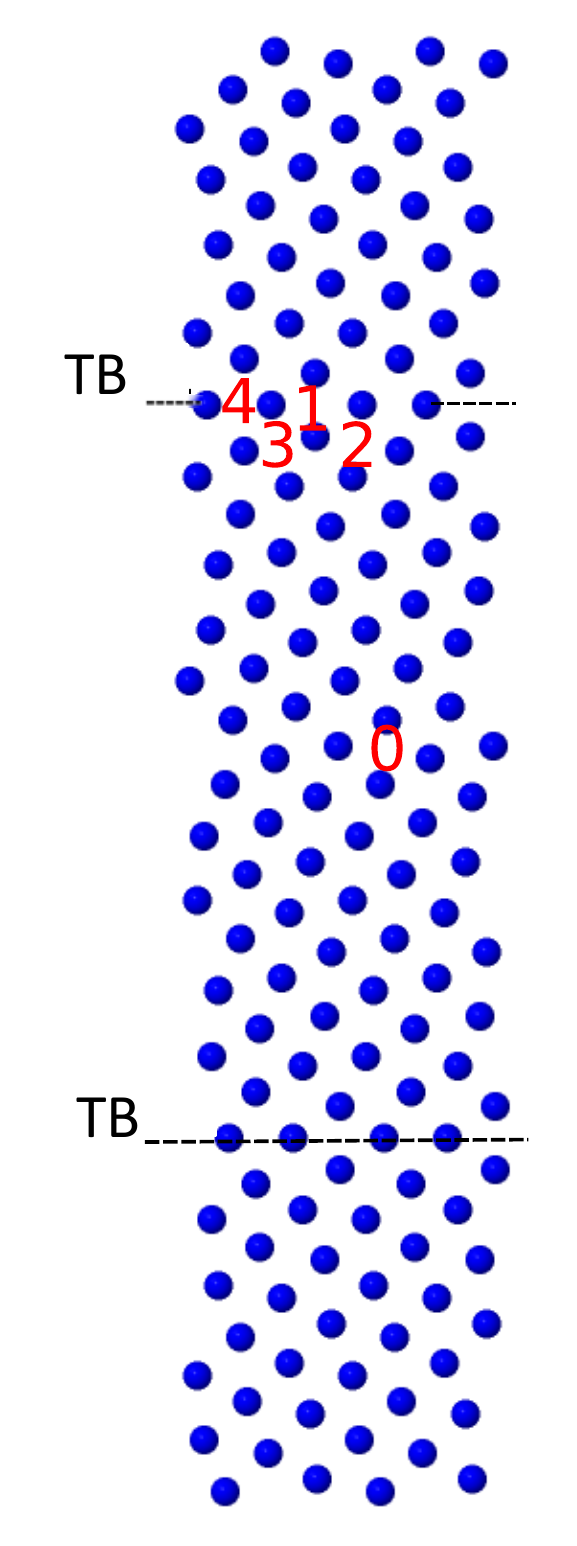}
 \end{center}

\caption{Oxygen interstitial sites around the twin boundary. Four sites, labeled TB-1 through 4, in the $\TB$ twin geometry are considered and compared with the octahedral site in bulk, 0. Interstitials on and below the interface are chosen. The top half of the twin boundary is related to the bottom half by mirror symmetry. Bulk geometry is retrieved away from the interface. To study the effect of periodic images of oxygen interstitial, three supercell dimensions $a$, $2a$ and $3a$ in $1/3[1\bar210]$ direction (out of the plane) are considered where $a$=2.948\AA\ is the lattice parameter of Ti. }

\label{fig:sites}
\end{figure}

\fig{sites} shows the supercell for modeling the $\TB$ twin boundary with oxygen. The initial structure is constructed from the perfect Ti lattice. Then, the twin boundary is generated by applying a mirror symmetry operation. In order for periodic boundary conditions to be valid in all directions, including the direction normal to the boundary, two twin boundaries are created in the structure. All atom positions and the supercell dimension perpendicular to the twin boundaries are allowed to relax till the forces are smaller than 5 meV/\AA. Previous studies have shown that relaxing these degrees of freedom gives the accurate energies\cite{ref:yoo}. Bulk geometry is retrieved away from the interface and lattice structures on opposite sides of the boundary are related through mirror symmetry. We choose four interstitial sites, labeled TB-1 through TB-4 (c.f, \fig{sites}) in and around the twin boundary. The site 0 corresponds to the octahedral site for interstitials in the bulk lattice. To study the effect of periodic images of the oxygen interstitial, we consider three out-of-plane dimensions ($a, 2a, 3a$) for the supercell along $1/3[1\bar210]$ where a=2.948~\AA~is the Ti lattice parameter.~\tab{deltaE} shows the energy difference between putting oxygen at twin boundary interstitial sites TB-1 through TB-4 and the bulk octahedral site 0. At each site \textit{i}, energy difference $\Delta E(\text{TB-}i)$ is defined as $E(\text{TB-}i)-E(0)$. For all supercell dimensions, we find that TB-1 and TB-4 located exactly at the boundary are more attractive for the oxygen than the bulk octahedral site while TB-2 and TB-3 are repulsive.
\begin{table}[!ht]
\caption{Oxygen energy at twin boundary interstitial sites. Table entries show the interaction energy at sites TB-$i$ defined as $ \Delta E(\text{TB-}i)=E(\text{TB-} i)-E(0)$. The energies are in meV units. For all supercell dimensions, TB-1 and TB-4 located exactly at the boundary are more attractive for the oxygen than the bulk octahedral site while TB-2 and TB-3 are repulsive.}
\label{tab:deltaE}

\centering
\begin{ruledtabular}
\begin{tabular}{cccc}
 & \multicolumn{3}{c}{ Dimension along $1/3[10\bar12]$} \\
 & $a$ & $2a$ & $3a$\\ 
\hline 
TB-1& --206.7 & --52.3 & --73.3\\
TB-2 & 48.6 & 114.6 & 105.9 \\
TB-3 & 57.2 & 192.8 & 176.7 \\
TB-4 & --112.7&--78.8 & --79.3
\end{tabular}
\end{ruledtabular}
\end{table}

\begin{figure}[!ht]
 \begin{center}
   \includegraphics[width=6 in]{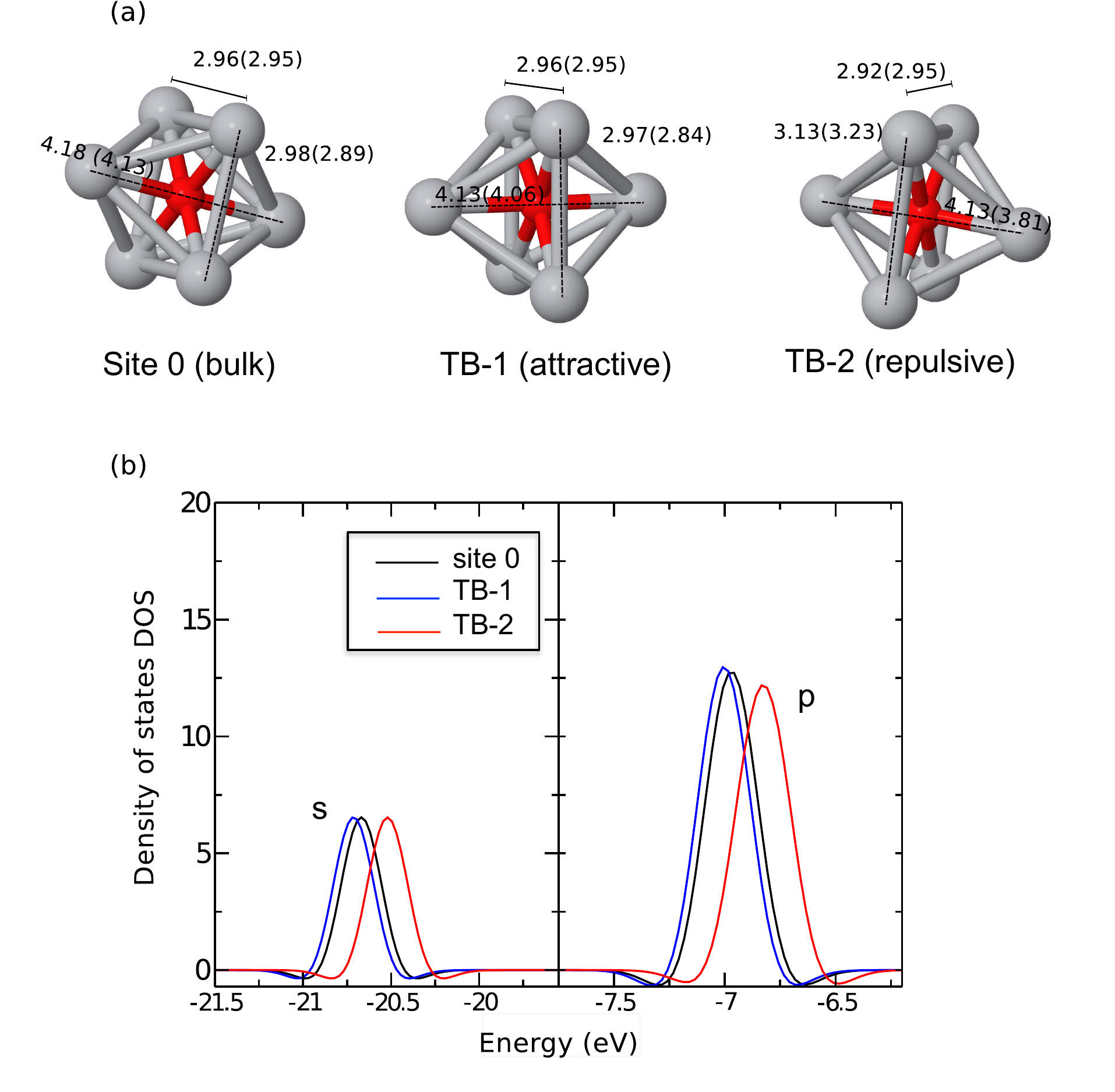}
 \end{center}

\caption{ (a) Geometric analysis of bulk reference site 0, attractive TB-1 and repulsive TB-2 interstitial sites and (b) electronic density of states for oxygen \textit{s} and \textit{p} states at each site. Site 0 is the bulk octahedral site, and shows the relaxation in the ground state for oxygen. (a) shows atoms at both TB-1 (attractive) and TB-2 (repulsive) sites undergo relaxations to approach the bulk geometry. The numbers in parenthesis show distances before relaxation, with distances are in \AA. In (b), the reference energy for the density of states is the Fermi energy. Attractive and repulsive sites show shifts in the oxygen states, but not changes in broadening.}

\label{fig:dos}
\end{figure}
\fig{dos}a shows the geometry of three interstitial sites with the local relaxations around the oxygen atom. We selected an attractive site (TB-1), a repulsive site (TB-2) and the bulk-equivalent site 0 for comparison. Examination of the atomic positions before and after relaxation reveals that atoms around both TB-1 (attractive) and TB-2 (repulsive) sites move to approach 0 geometry. The atomic distances are compared in \fig{dos}a. In addition, the electronic density of oxygen \textit{s} and \textit{p} states is shown in \fig{dos}b. Attractive and repulsive sites show shifts in the oxygen density of states, without any changes in bonding of the oxygen atom at the three selected sites. Different trends in interaction energy are likely to be caused by changes in elastic interactions due to relaxation of Ti atoms around the oxygen interstitial.

In addition, we determine the oxygen-oxygen pair interactions at each site by comparing data from different supercells where periodic images of oxygen are located at distances $a, 2a$ and $3a$. \tab{ebind} shows the change in oxygen binding energy as the distance between oxygen atoms changes. At each site, the difference in binding energy as oxygen-oxygen distance goes from $d_1$ to $d_2$ is defined as
\begin{equation}
\Delta E^b(d_1,d_2) = \left[E(S_{d_2}+\text{O})-E(S_{d_2})\right]-\left[E(S_{d_1}+\text{O})-E(S_{d_1})\right]
\end{equation}
where $S_{d_1}$ and $S_{d_2}$ denote the supercells with dimensions $d_1$ and $d_2$ respectively and $E(S_{d_i}+\text{O})$ is the energy of supercell $S_{d_i}$ with an oxygen at the corresponding interstitial sites. At the bulk site, the oxygen binding energy decreases when the distance between oxygen atoms doubles. This indicates that there is a repulsive oxygen-oxygen interaction at distance $a$. Increasing the distance to $2a$ does not affect the binding energy significantly; hence, the oxygen-oxygen interaction is weak at a distance of $2a$. This is consistent with a previous study of oxygen-oxygen pair interaction in Ti~\cite{ref:tio,Burton:2012uq}. At TB-1, increasing the separation of oxygen pairs from $a$ to $2a$ costs 37 meV, indicating an attractive interaction between oxygen pairs at distance $a$. The energy decreases when moving the atoms from $2a$ to $3a$ suggesting that the interaction is repulsive at $2a$ although the change in binding energy of oxygen pairs is still positive. At TB-2, oxygen pair interactions are repulsive at distance $a$ and attractive at $2a$. 

\begin{table}[!ht]
\caption{Change in oxygen binding energy with oxygen-oxygen distance. $\Delta E^b(a,2a)$ is the difference in binding energy as the oxygen-oxygen separation goes from a to 2a; a negative number indicates that the shorter distance is \textit{repulsive}, while positive values indicate \textit{attractive} binding between oxygen atoms at the shorter distance. The energy values are expressed in meV.}
\label{tab:ebind}

\centering
\begin{ruledtabular}
\begin{tabular}{ccc}
         & $\Delta E^b(a,2a)$ &$\Delta E^b(a,3a)$ \\ 
\hline
bulk & --117.7 & --116.7 \\
TB-1 & 37 & 16.5 \\
TB-2 & --49.1 & 19.5 
\end{tabular}
\end{ruledtabular}
\end{table}

\section{stacking fault}
\begin{figure}[!ht]
 \begin{center}
   \includegraphics[width=3 in]{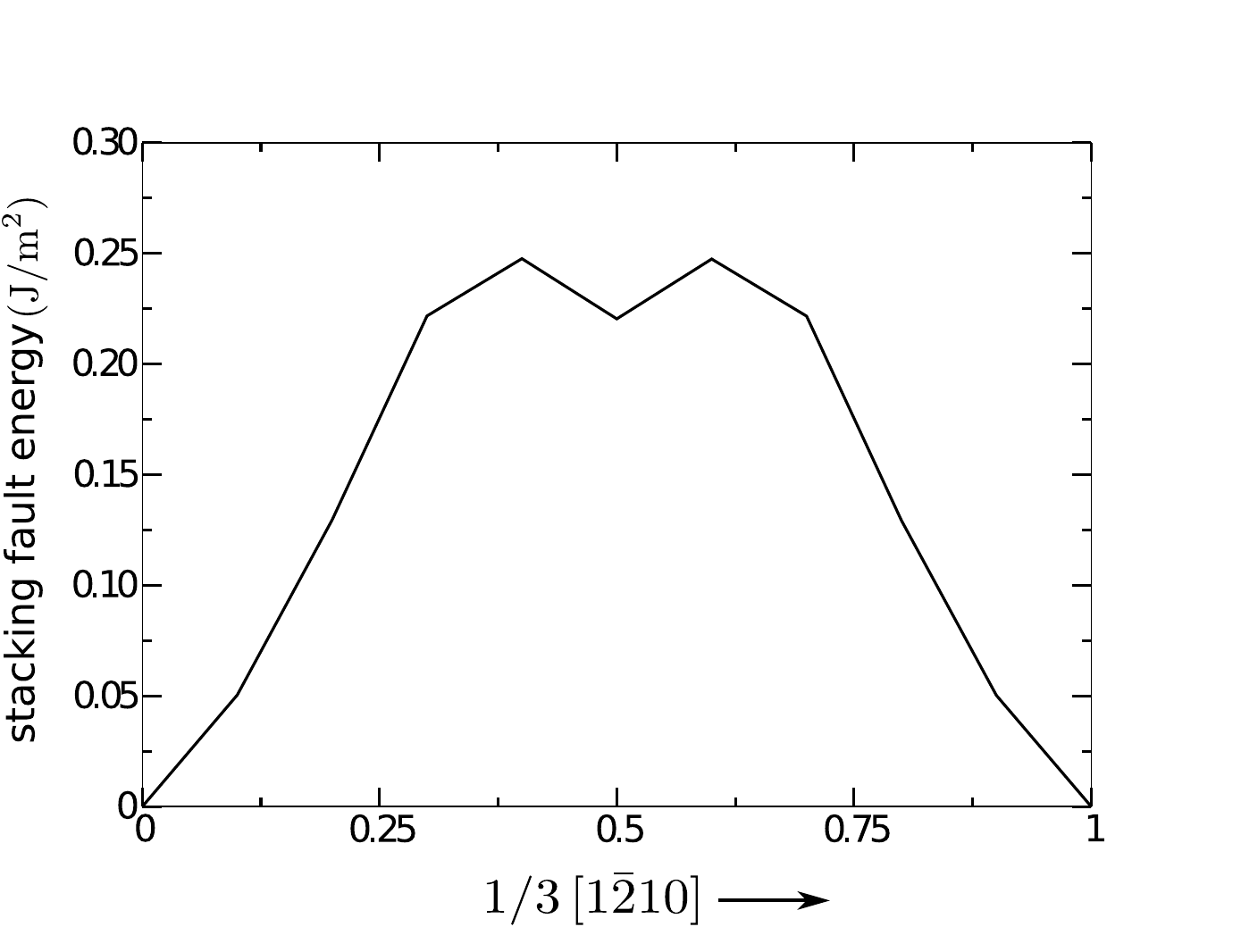}
 \end{center}

\caption{Generalized stacking fault energy for the $\SF$ prismatic plane along $1/3[1\bar210]$. DFT calculations give a metastable stacking fault at $a/6[1\bar210]$.}

\label{fig:plot}
\end{figure}

\fig{plot} shows the generalized stacking fault energy of the $\SF$ prismatic plane along $1/3[1\bar210]$. Prismatic slip is the dominant mode in Ti, and prismatic stacking faults are instrumental in studying core structure of dislocations in Ti.\cite{ref:screw} DFT calculations give a metastable stacking fault at the halfway point $a/6[1\bar210]$, with an energy of 0.220 J/$\text{m}^2$. This value is well established from other DFT calculations\cite{ref:yoo}. This fault has an interesting geometry, as it is not related to other closed-packed structures like the intrinsic basal faults; moreover, it is not expected to be low energy from a purely geometric consideration, but rather is dependent on the $d$-state bonding.

\begin{figure}[!ht]
 \begin{center}
   \includegraphics[width=6 in]{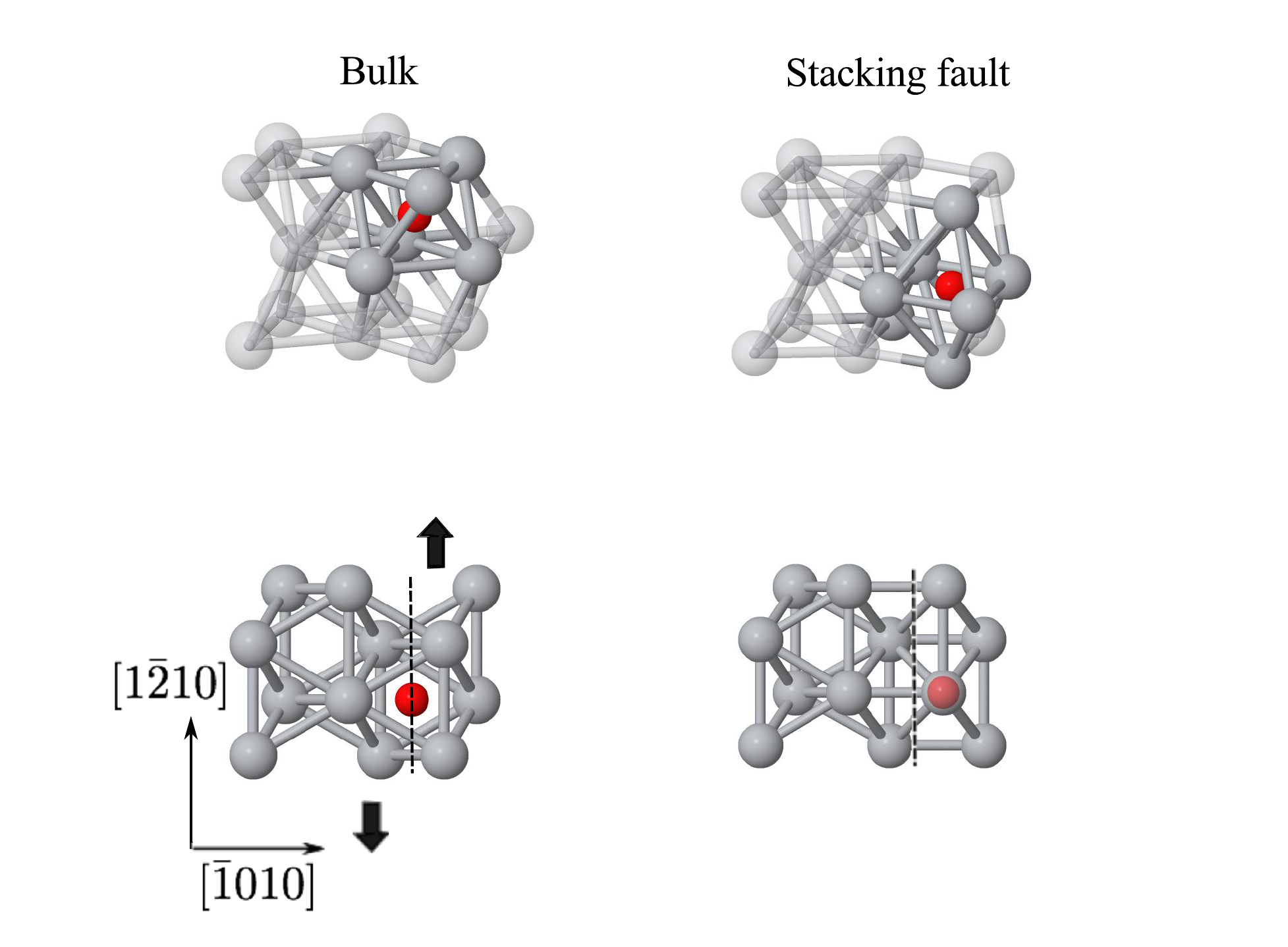}
 \end{center}

\caption{Geometry of an oxygen interstitial in the prismatic stacking fault. Structure of the faulted region locally changes from hcp to bcc. Oxygen moves from bulk hcp octahedral site to the  bcc octahedral site and increases the prismatic stacking fault energy.}

\label{fig:sf}
\end{figure}

\fig{sf} shows the geometry of an oxygen interstitial in the prismatic stacking fault before and after the $a/6[1\bar210]$ displacement. The local geometry for the octahedral interstitial site changes from hcp octahedral (6 neighbors forming a near-ideal octahedron) to a bcc-like octahedral (square plane of four close neighbors with two far out-of-plane neighbors). Our calculations show that oxygen---initially at the bulk hcp octahedral site---moves to the faulted region bcc octahedral site as shown in \fig{sf}. Since oxygen is an $\alpha$ stabilizer, it is expected to have higher energy in the bcc structure and therefore should increase the stacking fault energy~\cite{ref:conrad}. 
To quantify the effect of oxygen on the prismatic stacking fault, we used 10$\times$1$\times$2 (40 atoms) and 10$\times$2$\times$2 (80 atoms) Ti supercells with one and two lattice vectors dimensions along $[1\bar210]$ respectively. Using two sizes provides an understanding of the size effect resulted from having periodic images of the oxygen interstitial. \tab{sf} compares the stacking fault energies before and after introducing the oxygen interstitial to each supercell. Oxygen increases the stacking fault energy as expected. In addition doubling the supercell dimension results in a 6\%\ error. We also studied the oxygen-oxygen pair interaction at the stacking fault following the method we used for the twin boundary. The oxygen binding energy is defined as
\begin{equation}
E_\text{bind}(O)=\left[E^\text{fault}(\text{Ti}+\text{O})-E(\text{Ti}+\text{O})\right]-\left[E^\text{fault}(\text{Ti})-E(\text{Ti})\right].
\end{equation}
\tab{sf} shows that the oxygen binding energy increases as the oxygen-oxygen distance doubles leading to the fact that oxygen pair interactions at distance $a$ is attractive. This is a surprising difference with the bulk interaction, where there is a strong first-neighbor repulsion.

\begin{table}[!ht]
\caption{Effect of oxygen on the stacking fault energy. The prismatic stacking fault energy at 50\%\ along $[1\bar210]$ is calculated for pure Ti and two Ti supercells with an oxygen interstitial. $E_\text{bind}(O)$ is equal to $\left[E^\text{fault}(\text{Ti}+\text{O})-E(\text{Ti}+\text{O})\right]-\left[E^\text{fault}(\text{Ti})-E(\text{Ti})\right]$}

\label{tab:sf}

\centering
\begin{ruledtabular}
\begin{tabular}{ccc}
         & SF energy $(\text{J}/\text{m}^2)$ & $E_\text{bind}(O)$ (meV)\\ 
\hline
pure Ti & 0.220 & --- \\
40 Ti+O & 0.278 & 98.84\\
80 Ti+O & 0.262 & 145.28
\end{tabular}
\end{ruledtabular}
\end{table}

\section{conclusions}
We studied the energetics of oxygen interstitial interactions with $\TB$ twin boundary and a prismatic $\SF$ stacking fault. Four interstitial sites in the twin geometry are compared with the bulk octahedral site. We show that two of these sites located at the twin boundary are more attractive to oxygen than bulk while the sites away from the boundary are repulsive. In addition, we compared the oxygen-oxygen pair interactions at each site for different distances between periodic images of the oxygen atom. We observed that oxygen pairs repel/attract each other when separated by a single lattice parameter at the bulk/twin boundary site. Moreover, we study the interaction of oxygen with the prismatic stacking fault to approximate oxygen-dislocation interaction. We show that oxygen increases the stacking fault energy and therefore is repelled by the faulted geometry and consequently a dislocation core. Our results provide various pieces of quantitative information on the interaction of oxygen interstitials with defects in Ti; oxygen/twin boundary interaction energies reveal the existence of attractive sites at the twin boundary. This highlights the fact that atomic-scale features of oxygen/Ti interaction, which is lost in purely crystallographic models, are essential and should be considered. Moreover, the difference in first-neighbor interaction between oxygen atoms suggests that the repulsion is not a simple elastic repulsion, but depends strongly on the local configuration of neighboring titanium atoms. To obtain the complete picture of thermodynamics and kinetics of oxygen around the twin boundary, further calculations of the energy barriers for oxygen diffusion between various interstitial sites around the twin boundary should be performed. Moreover, computation of the interaction energy between pairs of oxygen atoms at various separations are essential in any oxygen/defect simulations involving periodic boundary conditions. Based on these calculations we conclude that direct DFT simulations of dislocation/oxygen interactions require at least $3a$ periodic lengths along the dislocation threading direction. This work is a first step towards quantifying the effect of oxygen on mechanical properties of Ti.

\begin{acknowledgments}
This research was supported by NSF/CMMI CAREER award 0846624, Boeing, and by the Office of Naval Research through ONR Award No. N000141210752. The authors gratefully acknowledge use of the Turing and Taub clusters maintained and operated by the Computational Science and Engineering Program at the University of Illinois; as well as the Texas Advanced Computing Center (TACC) at the University of Texas at Austin.
\end{acknowledgments}

%

\end{document}